# Solid-state-processing of δ-PVDF

Jaime Martín[1], Dong Zhao[2] (赵冬), Thomas Lenz[2,3], Ilias Katsouras[4], Dago M. de Leeuw[2,5] and Natalie Stingelin[1,6]*

[1] *Department of Materials and Centre of Plastic Electronics, Imperial College London, London SW7 2AZ, UK*
[2] *Max-Planck Institute for Polymer Research, Ackermannweg 10, 55128 Mainz, Germany*
[3] *Graduate School Materials Science in Mainz, Staudinger Weg 9, 55128 Mainz, Germany*
[4] *Holst Centre, High Tech Campus 31, 5656AE Eindhoven, The Netherlands*
[5] *Faculty of Aerospace Engineering, Delft University of Technology, Kluyverweg 1, 2629 HS Delft, The Netherlands*
[6] *School of Materials Science and Engineering and School of Chemical & Biomolecular Engineering, Georgia Institute of Technology, 311 Ferst Drive, Atlanta, Georgia 30332, USA*

[*] E-mail: natalie.stingelin@mse.gatech.edu

Keywords: δ-PVDF, piezoelectricity, ferroelectricity, solid-state-processing, polymorphisms.

**Poly(vinylidene fluoride) (PVDF) has long been regarded as an ideal piezoelectric 'plastic' because it exhibits a large piezoelectric response and a high thermal stability. However, the realization of piezoelectric PVDF elements has proven to be problematic, amongst others, due to the lack of industrially-scalable methods to process PVDF into the appropriate polar crystalline forms. Here, we show that fully piezoelectric PVDF films can be produced via a single-step process that exploits the fact that PVDF can be molded at temperatures below its melting temperature, *i.e.* via solid-state-processing. We demonstrate that we thereby produce δ-PVDF, the piezoelectric charge coefficient of which is comparable to that of biaxially stretched β-PVDF. We expect that the simplicity and scalability of solid-state processing combined with the excellent piezoelectric properties of our PVDF structures will provide new opportunities for this commodity polymer and will open a range of possibilities for future, large-scale, industrial production of plastic piezoelectric films.**

Since the discovery in the 1970s that poly(vinylidene fluoride) (PVDF) [1,2] can display a large piezoelectric response, this plastic has been the subject of intense research, and ubiquitous applications in everyday life have been foreseen, for instance, in sensing, energy harvesting, acoustics and information storage [3,4,5,6,7]. However, processing PVDF into a piezo-/ferroelectric form has proven to be challenging, amongst others due to the lack of industrially-scalable methods that can be applied for this purpose. The reason for this difficulty is that the ferro- and piezoelectricity of PVDF is linked to the coherent spatial distribution of the local C-F dipoles along the polymer backbone and its macromolecules' long-range packing order. Because of the variety of chain- and dipole arrangements that can be adopted, PVDF indeed exhibits at least four well-identified polymorphs, referred to as the *α*, *β*, *γ* and *δ*-phase [8,9]. The *α*-phase is at ambient conditions the thermodynamically stable polymorph. Fig. 1a shows the projection of the chain arrangement in *α*-PVDF along the *c* axis of the unit cell. The orthorhombic, centrosymmetric unit cell contains two chains in $tg^+tg^-$ conformation [10,11]. Due to the anti-parallel packing of the chains in the unit cell, their dipole moments cancel out, rendering the *α*-phase non-polar and paraelectric. In contrast, the other PVDF polymorphs, *i.e.* the *β*-, *γ*- and *δ*-phases, are polar, hence ferroelectric and concomitantly piezoelectric [9]. The *β*-PVDF is the most pursued phase because it exhibits the highest ferroelectric and piezoelectric properties amongst the PVDF polymorphs — and of all polymers in general [2]. The orthorhombic unit cell has two chains in an all-*trans (ttt)* conformation [12] resulting in a net dipole moment per unit cell of $8 \times 10^{-30}$ Cm [13]. Since PVDF is a semicrystalline polymer with a crystallinity of about 50%, it can be estimated that thin films made of this PVDF polymorph should feature a remanent polarization of around 9 μC cm$^{-2}$. The *γ*-phase of PVDF, which can be regarded as a mixture of the *α*- and *β*-phase, crystallizes in a monoclinic unit cell with four chains in a $t_3g^+t_3g^-$ conformation [14]. Therefore, the dipole moment is smaller than that of the *β*-phase. The third polar phase, the so-called *δ*-phase, is known since 1978 [9,15,16,17,18]. The crystal structure of this forgotten phase has recently been refined [5,19]. The *δ*-phase is the polar version of the *α*-phase; both phases have the same lattice constants and chain conformation ($tg^+tg^-$) [15,19]. However, in *δ*-PVDF, every second chain is rotated 180º around the chain axis (Fig. 1a). In order to prevent impossibly small interchain fluorine-fluorine distances, the macromolecules are shifted by half of the *c*-axis lattice constant. As a consequence, the remanent polarization, *i.e.* the displaced charge density at zero bias, amounts to 7 μC cm$^{-2}$ and, thus, is comparable to that of *β*-PVDF.

Crystallization of PVDF from the melt or solution typically leads to the non-polar, paraelectric *α*-phase. Reported processing routes to obtain one of the polar PVDF phases are summarized in Fig. 1b. *β*-PVDF is commercially available, but it is produced in a two-step-process by biaxially stretching melt-processed α-PVDF, limiting the product range to free-standing foils. γ-PVDF is experimentally very challenging to access although recent work indicates that it can be formed in spatially confined systems [20, 21]. Finally, δ-PVDF has only been realized by applying high electric fields (≥ 170 MV/m) on α-PVDF structures. This process is called electroforming and it frequently results in the breakdown of both the electrode and the polymer [19] – an issue that has rendered δ-PVDF so far an impractical option from an industrial perspective.

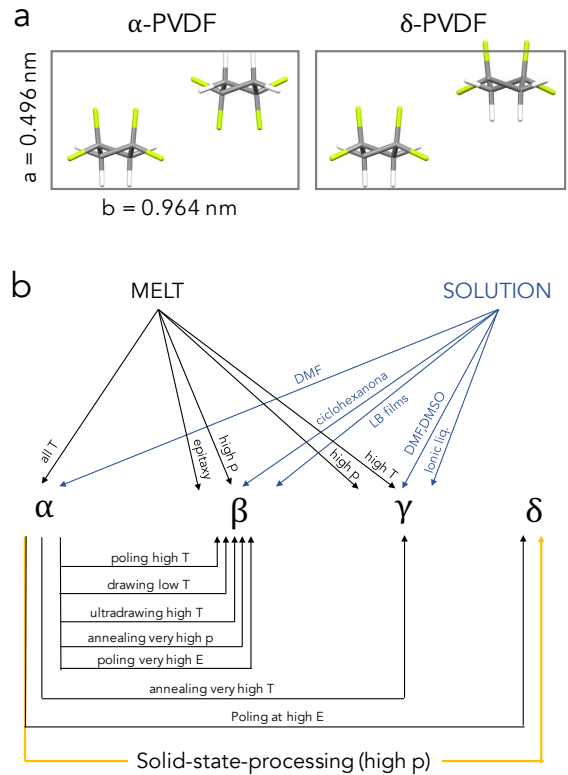

**Fig. 1**. **Molecular arrangement and processing routes towards polar PVDF phases** (adapted from [9]). a) Projections onto the *a-b* plane of the unit cell of the chain arrangement in, respectively, α- and δ-PVDF. (b) Processing routes of PVDF towards polar forms. δ-PVDF has been hitherto realized by applying high electric fields (≥ 170 MV/m) on α-PVDF, which frequently results in the breakdown of both the electrode and the polymer. Solid-state processing provides a simple, cost-efficient route to produce δ-PVDF films.

Here, we present a simple route to produce piezoelectric δ-PVDF films allowing the assessment of the piezoelectric charge coefficient, $d_{33}$, of this PVDF polymorph for the first time after almost 40 years [16, 22]. Our method exploits the solid-state-processing of α-PVDF, *i.e.* the application of moderate pressures (~20 kN/cm$^2$) below the melting temperature, to induce a phase transition from the paraelectric α-PVDF phase to the polar δ-phase. Solid-state processing is already used, for instance, to manufacture products of poly(tetrafluoroethylene) (PTFE – better known as Teflon® [23]). The versatility of this process has recently been further highlighted by the successful fabrication of organic semiconducting films with high bulk charge-carrier mobility [24]. In the case of PVDF, solid-state processing is efficient and simple; indeed, it leads in one single step to δ-PVDF films – in contrast to the relatively elaborate mechanically stretching of α-PVDF films to produce foils of β-PVDF currently employed in industry.

Films of PVDF were prepared by compression molding commercially purchased powder in a hot press at temperatures below the melting temperature of α-PVDF, *i.e.* at temperatures between 140 ºC and 150 ºC. The pressure was thereby progressively increased up to ~20 kN/cm$^2$ and applied for 5 min. Subsequently, the films were cooled to ambient temperature under pressure. No coherent films could be produced when the pressing temperature was below 130 ºC; and when pressing temperatures between 130ºC and 140 ºC were used, films were obtained that were generally heavily cracked. In strong contrast, at higher pressing temperatures, the process led to highly transparent films of ~30 μm thickness and an area of about 1 cm$^2$.

In order to assess the ferroelectric properties of such solid-state pressed films (pressed at temperatures above 140 ºC), we measured hysteresis loops after evaporating 30 nm Au electrodes on both sides of the free-standing structures (Fig. 2). In the first set of measurements, with increasing amplitude of the applied electric field from 30 MV/m to 250 MV/m, the loops gradually open (Fig. 2a). The inner loops only slightly exceed the outer loops, which indicates that electroforming can be disregarded as a source of the piezoelectric response. We emphasize that the opening of the inner loops was gradual and not step-wise. The latter occurs in the case electroforming plays a role. Clearly, the as-prepared, solid-state films are ferroelectric, implying that this process directly induces a polar PVDF phase, which displays a coercive field, $E_c$, of 110 MV/m for the saturated loop, and a saturated polarization of 7 μC cm$^{-2}$.

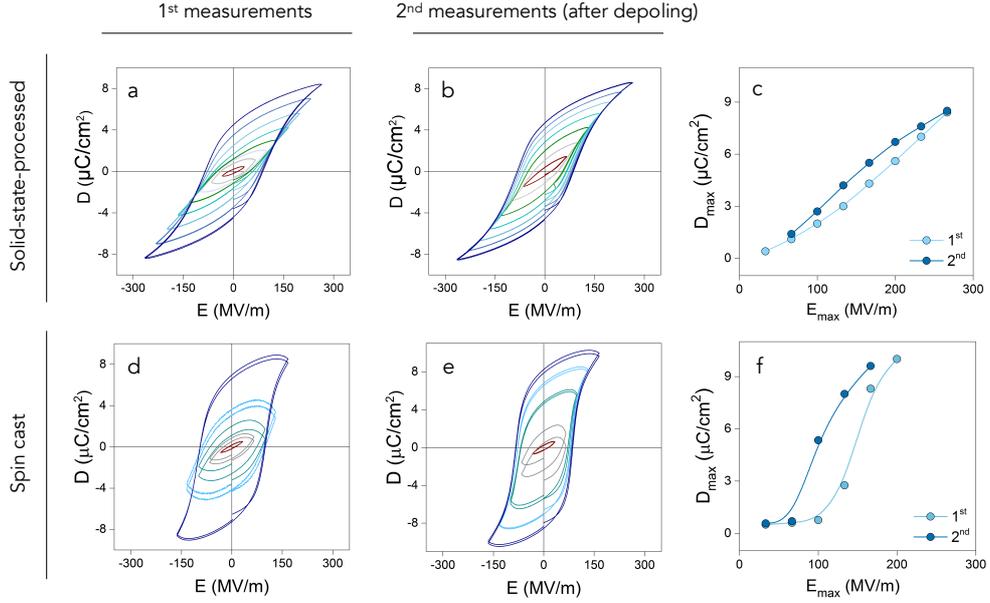

**Fig. 2 Ferroelectric hysteresis loops of solid-state pressed and spin-cast PVDF films**. (a) Hysteresis loops of solid-state processed PVDF films measured with stepwise increasing electric field (amplitude). (b) Re-measured hysteresis loops after electrically depoling the film (a). (c) Maximum displacement as a function of electric field for a solid-state processed PVDF film as obtained from the first and second set of measurements. (d) Hysteresis loops of a spin-coated PVDF film measured with stepwise increasing electric field (amplitude) which shows that the initially prepared film is electroformed into a polar PVDF at high electric field. (e) Re-measured hysteresis loops after electroforming the spin-cast PVDF film. (f) Maximum displacement as a function of electric field for a spin-coated PVDF film as obtained from the first and second set of measurements.

We repeated the hysteresis measurements after electrically depolarizing the solid-state pressed films with an alternating electric field of decreasing amplitude [5] — similar to the procedures used to demagnetize a ferromagnetic material [25]. We find that the second set of hysteresis loops (Fig. 2b) are similar to the loops obtained in the first measurements, which further confirms that solid-state processing directly leads to a polar, ferroelectric PVDF film.

For comparison, we performed control measurements on free-standing, spin-cast PVDF films, which consist of the $\alpha$-phase. The first set of measurements (Fig. 2d) show that at low electric fields, the loops resemble those obtained for a leaky capacitor. However, when the electric field exceeded 170 MV/m, the loops abruptly open, adopting a hysteretic shape typical for a ferroelectric material. This stepwise behavior indicates that at high electric field, spin-cast films are electroformed into a polar PVDF phase. In accord with this observation is the fact that the loops obtained in the second measurements (Fig. 2e) gradually open, resembling those of the solid-state pressed films.

The formation of polar PVDF upon hot pressing can also be confirmed by plotting the extracted values of the maximum value of the displacement as a function of electric field. Fig. 2c shows that for solid-state processed PVDF the values for the first and second measurement are very similar, establishing that polar PVDF was formed in one step during compression molding. In contrast, for spin-cast films the values for the first and the second set of measurements differ strongly (Fig. 2f). This stepwise increase of displacement with electric field implies that α-PVDF is electroformed into polar PVDF, leading after electroforming to a relation between maximum displacement and electric field that is similar to that of solid-state processed polar PVDF.

Having established that a polar, piezoelectric PVDF phase is formed via solid-state processing, we measured simultaneously the electric displacement and strain of the solid-state pressed films. The displacement as a function of electric field is presented in Fig. 3a, and the strain as function of electric field and of displacement is presented in Fig. 3b,c. The strain can be quantitatively described by a model previously used to explain the negative piezoelectric effect of PVDF and its copolymers [2]. The strain comprises the polarization-induced electrostrictive strain and an additional term resulting from the electro-mechanical coupling between the crystalline and amorphous fractions of PVDF. The displacement and strain can be simultaneously fitted as a function of electric field. Intriguingly, a perfect fit is obtained for the solid-state pressed PVDF films (Fig. 3). From the fit, we can extract a value of the piezoelectric charge coefficient, $d_{33}$, of -36 pm/V. Interestingly, this value is comparable to that of biaxially stretched β-PVDF (-31 pm/V), and it is slightly higher than that of electroformed δ-PVDF (respectively, -13 pm/V [16] and -15 pm/V [22]).

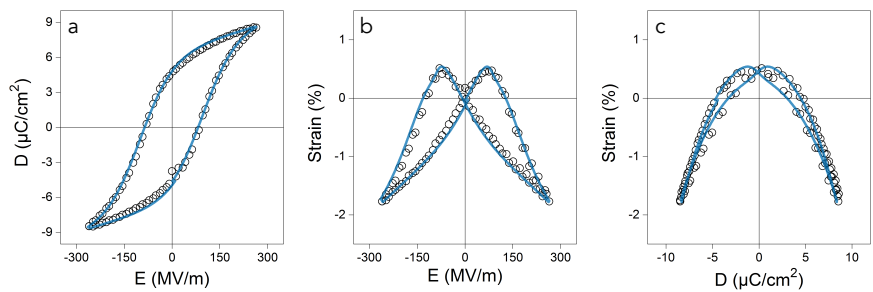

**Fig. 3 Ferroelectric hysteresis and piezoelectric strain in solid-state pressed PVDF films.** Measured (black) and modelled (red) electric displacement curves (a) and strain (b), both measured *vs.* electric field. (c) Corresponding strain *vs.* electric displacement loops.

The question that remains to be answered is what polar PVDF phase is induced during the solid-state pressing process. To address this issue, we conducted wide-angle X-ray scattering (WAXS) measurements. The WAXS patterns of as-received PVDF powder (light blue) and a solid-state pressed film (dark blue) are displayed in Fig. 4a. From the peak positions, we can unambiguously deduce that the purchased powder is $\alpha$-PVDF, in agreement with literature. Since (i) both samples (as-received powder and solid-state pressed film) feature identical X-ray diffraction patterns (Fig. 4a and Supporting Information S2), (ii) it is known that the polar $\delta$-PVDF features the same crystal structure as the non-polar $\alpha$-phase, and (iii) the solid-state pressed films are piezoelectric, we further conclude that the solid-state pressed sample is comprised of the $\delta$-phase.

The similarity in the X-ray diffraction pattern of the $\delta$-phase with $\alpha$-PVDF renders, however, full identification of the $\delta$-phase by X-ray diffraction challenging[19]. Some conclusions can nonetheless be drawn. For instance, the shoulder at ~21° that we observe for the solid-state pressed films indicates that a small amount of $\beta$-PVDF is present in our solid-state pressed films. We like to note, though, that when deconvoluting these WAXS pattern into the contributions of the $\alpha$, $\beta$ and $\delta$ phases (Fig. 4b), we deduce that the amount of $\beta$-phase in the films is less than ~10 %, which is far too small to dominate their piezoelectric behavior.

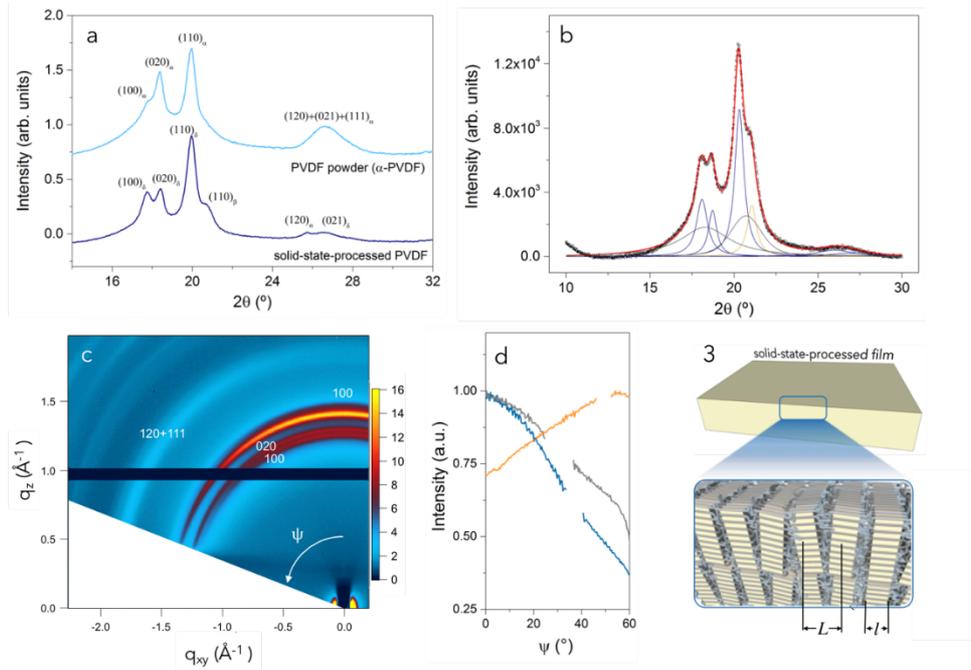

**Fig. 4. X-ray diffraction data of solid-state-processed PVDF films.** (a) WAXS $\theta$-$2\theta$ scans of solid-state-processed PVDF (dark blue line) and commercially bought PVDF powder (*i.e.*, $\alpha$-PVDF; light blue line). (b) Decomposition of the WAXS $\theta$-$2\theta$ scans into elementary peaks of the $\delta$-, $\alpha$-, and $\beta$- phases. Relevant diffractions of the $\delta$- and $\alpha$- phase are plotted in blue, while those of the $\beta$-phase are shown in orange and grey. The broad peaks correspond to the contribution of the amorphous halo. (c) 2D-WAXS pattern of solid-state processed PVDF films ($z$ corresponds to the direction normal to the film $x$-$y$ plane). (d) WAXS intensity of the (110) reflection (blue line), ((100) + (020)) reflections (grey line) and ((120) + (021)) reflections (orange line) plotted against the $\psi$-angle, which is the angle defined by the rotation of the sample around an axis contained in the plane of the film ($x$-$y$ plane). (e) Schematic illustration of the preferential crystal orientation that we deduce from our X-ray data presented in a,b for solid-state-processed films.

We also conducted 2D-WAXS experiments to gain further insights into solid-state processed PVDF films. Fig. 4c shows the diffracted WAXS intensity (in colour scale) of solid-state pressed films, plotted against $q_z$ and $q_{xy}$, *i.e.* the wave vectors normal and parallel to the plane of the film. Arch-like diffractions were recorded for all Bragg maxima indicating a continuous distribution of crystal orientations within these films. The distribution can be obtained from the intensity profiles of the diffraction maxima along the $\psi$-angle, that is the angle defined by the rotation of the sample around an axis contained in the plane of the film ($x$-$y$ plane). Fig. 4d shows that the intensity of (110) and (100+020) reflections decrease when approaching the equator, while that of the (021) reflection increases. These tendencies are indicative of a preferred crystal orientation where the $c$-axis (*i.e.* the direction of along the chains) of this PVDF polymorph points parallel to the plane of the film, as schematically depicted in Fig. 4e (see also Supporting Information Fig. S3), similar to what was reported for poly(3-hexylthiophene) (P3HT) when it was processed via solid-state pressing [24].

Further information was obtained from estimating the lamellar crystal thickness, $l$, of the ordered moieties present in solid-state pressed structures, using for this purpose small-angle X-ray scattering and differential scanning calorimetry (DSC) data (see Supporting Information S4). We have indicated $l$ in the schematic of the solid-state processed structure displayed in Fig. 4e. Both solid-state-processed PVDF and α-PVDF films, produced from solution, feature well-defined Bragg peaks in the low $q$-region of the respective SAXS profiles (Supporting Information Fig. S4b). The peak for the solid-state pressed PVDF is centred at $q \sim 0.049$ Å$^{-1}$, which corresponds to a characteristic length of the periodic stack, *i.e.* the so called long period $L$, which is comprised of the thickness of the crystalline lamella as well as the one of the unordered fraction in this stack (see also Fig. 4e). We find that $L \approx 12.8$ nm for solid-state pressed films, while for $\alpha$-PVDF (processed from solution), a long period $L \approx 12.1$ nm is extracted.

Since the lamellar crystal thickness can be calculated from $l = L \cdot X$ where $X$ is the degree of crystallinity, we can deduce $l$ when estimating $X$ from thermal analysis data. Both, $\alpha$- and solid-state pressed PVDF display similar thermograms with a single dominant endotherm that is associated with the melting of the crystalline fraction in these structures (Supporting Information Fig. S4a). Assuming that both phases possess equal values for the enthalpy of fusion, $\Delta H_\mathrm{f}$, for a 100% crystalline material of 104.5 J/g [26], we estimate $X$ to be 42% for solid-state pressed PVDF and 44% for $\alpha$-PVDF. Clearly, and despite the high weight-average molecular weight of the PVDF used here (543 kg/mol), the degree of crystallinity does not change with solid-state-processing. The lamellar crystal thickness, $l$, is also similar in both cases and is calculated to be 5.4 nm.

The mechanism of the transition of $\alpha$-PVDF into $\delta$-PVDF remains unclear. The driving force for the change of molecular orientation upon high-field poling is probably the applied electric field, which induces the re-orientation of the normal component of the dipole moment of specific chain segments and, in turn, leads to the chain segments' physical rotation [27, 28]. However, in solid-state processing no electric field is applied; hence, the antiparallel-to-parallel chain reorientation that is required to induce a transition from the $\alpha$- to the $\delta$-polymorph, must have a different origin. We propose here that at the temperatures and pressures used during solid-state-processing, the $\delta$-PVDF may be the thermodynamically stable phase. The physical rotation of chain segments needed for the $\alpha$- to $\delta$-phase transition to occur, might, thereby, be assisted by the dynamics of the PVDF chain segments within the $\alpha$-crystals. Indeed, while the PVDF macromolecules in the $\alpha$-polymorph are fully frozen at room temperature [29], a local dynamic process, commonly referred to the as the $\alpha_c$ relaxation, starts to be operative above 100-120 °C [9, 29] — *i.e.* in the temperature range where solid-state pressing started to result in fully compact and optically transparent films. The precise molecular motion leading to this $\alpha_c$ relaxation is not fully understood; however, it is generally accepted that it originates from rotational or conformational motions along the PVDF chains [29, 30, 31]. Hence, in the case of solid-state processing PVDF, certain bonds and atoms along this polymer's chains can exhibit rotational dynamics in the temperature regime where the $\alpha_c$ relaxation sets in. These rotations may facilitate the transition from an antiparallel-to-parallel orientation of the PVDF chains and, as a consequence, the phase transition from the $\alpha$- to $\delta$-polymorph. This hypothesis is supported by the fact that we were not able to fabricate coherent films in the solid state at temperatures below the $\alpha_c$-relaxation temperature.

Note, finally, that it was previously suggested that a transition from $\delta$-PVDF to the less polar $\gamma$-phase may occur at temperatures slightly below the melting of these PVDF phases [32]. This would lead to a reduction in the piezoelectric performance of the free-standing films. We, therefore, assessed the thermal behavior of solid-state pressed PVDF films with differential scanning calorimetry (Supporting Information Fig. S4a). Independent of the heating rate used, the melting endotherm is the only noticeable feature we observed for solid-state pressed material in the temperature range analyzed. This suggests that no solid-solid phase transition occurs in such PVDF structures. This finding is important from a technological point of view, because a potential transition into a less polar phase would be accompanied by loss of remanent polarization and a decrease of the piezoelectric charge coefficient.

In summary, we have demonstrated a one-step process that results in piezoelectric and ferroelectric PVDF films. Our method is based on compression molding commercially available PVDF powder in the solid state. We show that under moderate pressures and temperatures, a solid-state transition from the technologically relatively uninteresting apolar $\alpha$-phase to the polar, fully piezoelectric $\delta$-phase occurs. This transition may result from a higher thermodynamic stability of the $\delta$-phase under those conditions; it may also be assisted by the dynamics of the chain-segments within $\alpha$- PVDF, *i.e.* the $\alpha_c$ relaxation. The coercive field for the saturated ferroelectric loop for such solid-state pressed PVDF films is 110 MV/m, while the saturated polarization is 7 $\mu$C cm$^{-2}$. The piezoelectric charge coefficient is -36 pm/V. Intriguingly, this value is comparable to that of biaxially stretched $\beta$-PVDF (-31 pm/V); and it is notably higher than values previously reported for electroformed $\delta$-PVDF (-31 resp. -15 pm/V). The structural analysis of solid-state pressed PVDF reveals a partial orientation of the polymer chains in plane of the films, which likely is due to the compression forces that act during solid-state pressing. The degree of crystallinity of such structures can be deduced to be ~42%, while the lamellar thickness *l* was estimated to be 5.4 nm, with the long period *L* extracted as 12.8 nm. Since these values are similar to those obtained for α-PVDF, we conclude that solid-state-processing does not induce strong changes in the internal structure of the material other than the phase transformation to the δ-PVDF. Most importantly, the simplicity of solid-state processing to induce a polar, fully piezoelectric form of PVDF may open a range of new possibilities for this interesting material not only for further fundamental studies but also for future industrial production of piezoelectric plastic films.

## Methods

***Materials.*** Poly(vinylidene fluoride), PVDF, with a weight-averaged molecular weight, $\overline{M_w}$, of 543,000 g mol$^{-1}$, was purchased from Sigma-Aldrich Co. and used as received.

***Processing method.*** For solid-state-processing PVDF structures, as-received powder (10-20 mg) was placed in a hot press (Rondol Autopress), followed by compression molding at temperatures ranging from 100 to 150 ºC. Pictures of the material before and after processing are supplied in the Supporting Information (Fig. S1). During the compression molding process, the pressure was progressively increased up to ~10 to 30 kN/cm$^2$ and kept for a period that depended on the applied temperature. Typical conditions used were 150 ºC, ~20 kN/cm$^2$ and 5 min. The PVDF samples were then cooled down to room temperature under pressure to obtain ~30 μm thick films.

For small-angle X-ray measurements, PVDF films were produced via drop-casting from N,N-dimethylformamide (DMF).

Free-standing films of PVDF were in addition prepared by spin-coating for comparison. To that end, 50 nm films of PEDOT:PSS (Clevios P VP 4083, Heraeus) were spin-coated onto cleaned glass substrates (10 s at 500 rpm, followed by 60 s at 1500 rpm). Then, a PVDF solution (340 mg/ml in DMF) was spin-coated onto the PEDOT:PSS films at 250 rpm for five minutes, followed by the annealing of the samples at 200 °C for two hours. Samples were then slowly cooled down to room temperature and immersed in a water bath in order to detach the PVDF films from the substrate. The resulting free-standing films are by definition in the α-phase, as they were obtained by slow cooling from the melt (after spin-coating). They serve as reference for the electrical characterization.

***Characterization.*** The electric displacement as a function of electric field was measured using a Radiant precision multiferroic test system (Radiant Technologies, Inc.). The strain as a function of electric field was measured simultaneously with the displacement, using a MTI 2100 photonic sensor interfaced with the Radiant tester.

Wide-angle X-ray scattering (WAXS) θ/2θ scans were recorded at room temperature on a PANalytical X'Pert Pro MPD using Cu Kα radiation. The analyzed scattering vector, $q$ $(q = 4\pi/\lambda \sin\theta)$ was thereby parallel to the out-of-plane direction of the films. The amount of the different PVDF phases present in solid-state pressed structures was quantified by fitting the WAXS pattern using the *PeakFit4.12* software, taking as an assumption that the final crystallinity of the samples should be around 40 - 50 %. The peaks were fitted to Voight functions, while the amorphous halo was fitted as the sum of 2 peaks. One of the main reasons for the latter is the strong diffuse scattering in the region that is present between 15° - 20°, which renders description of the diffuse scattering of the amorphous PVDF fraction intricate. WAXS patterns for hypothetically defect-free *α*- and *δ*-PVDF were calculated using the Mercury 3.8 software (Supporting Information Fig. S2)

Two-dimensional wide-angle X-ray scattering (2D-WAXS) measurements were performed at the D-line of the Cornell High Energy Synchrotron Source (CHESS). The WAXS intensity was recorded along the $q_z$ and $q_{xy}$, *i.e.* the scattering vectors normal and parallel the plane of the film, respectively. A wide band-pass (1.47%) X-ray irradiation of a wavelength of 1.155Å was used, selecting an incidence angle between 0.5° and 1°. A Pilatus 200k detector with a pixel size of 172 μm was placed at a distance of 28.9 cm from the samples. A 1.5 mm wide tantalum rod was used to block the intense scattering in the small-angle area. The exposure time was 1s. The 2D-WAXS analysis was limited to angles $\psi$ (defined by the rotation of the sample around an axis contained in the plane of the film) between 0° and 65º due to the shadowing of a certain detector area by the sample. The (100) and the (020) reflections were computed together as their peaks overlapped.

Small-angle X-ray scattering (SAXS) experiments were conducted on a Rigaku 3-pinhole PSAXS-L equipment operating at 45 kV and 0.88 mA, employing Cu Kα radiations with a wavelength $\lambda = 1.54$ Å. The flight path and the sample chamber in this equipment were kept under vacuum. The scattered X-rays were detected on a two-dimensional multiwire X-ray detector (Gabriel design, 2D-200X). This gas-filled proportional type detector offers a 200 mm diameter active area with ca. 200 μm resolution. The azimuthally averaged scattered intensities were obtained as a function of $q$. Reciprocal space calibration was performed using silver behenate as standard. PVDF films were placed perpendicular to the incident X-ray beam and analyzed in transmission geometry. For the sake of comparison, α-PVDF films were also analyzed. These were processed from N,N-dimethylformamide (DMF) solution.

Differential scanning calorimetry (DSC) was performed on a Mettler–Toledo DSC1 Star system using heating- and cooling rates of 10 °C/min. 3-5 mg of freshly prepared solid-state-processed films were used for the analysis. We used the 1$^{st}$ heating thermogram to obtain information on the thermal behavior of solid-state-processed PVDF, while from the 2$^{nd}$ heating thermogram insights were gained on melt-processed material that leads to α-PVDF. Measurements were also conducted on as-received PVDF powder.

**Acknowledgements**

J.M. acknowledges support from the European Union's Horizon 2020 research and innovation programme under the Marie Skłodowska-Curie grant, agreement No 654682. T. L. acknowledges financial support by the Graduate School Materials Science in Mainz. N.S. is in addition grateful for support by a European Research Council ERC Starting Independent Research Fellowship under the grant agreement No. 279587. We further acknowledge A. Arbe and A. Iturrospe for their invaluable assistance with SAXS measurements, and L. Yu, R. Li and D.-M. Smilgies for 2D-WAXS experiments at CHESS.

**Author contributions**

N.S., D.M.d.L. and I.K. conceived the idea. J.M. prepared the solid-state processed films and performed the structural and thermal characterization and analyzed the data. T.L. and D.Z fabricated the spin-cast films and conducted the electrical characterization for all samples. D.Z. and I.K. analyzed the electrical results. J.M., D.Z., T.L., I.K., D.M.d.L., and N.S. co-wrote and commented on the manuscript. D.M.d.L. and N.S. supervised the project.

**Additional Information**

Supplementary information is available in the online version of the paper. Reprints and permissions information is available online at www.nature.com/reprints. Correspondence and requests for materials should be addressed to N.S.

**Competing financial Interests**

The authors declare no competing financial interests.

# References

1. Kawai H. The Piezoelectricity of Poly (vinylidene Fluoride). *Jpn. J. Appl. Phys.* **8,** 975 (1969).
2. Katsouras I., Asadi K., Li M., van Driel T. B., Kjaer K. S., Zhao D.*, et al.* The negative piezoelectric effect of the ferroelectric polymer poly(vinylidene fluoride). *Nat. Mater.* **15,** 78-84 (2016).
3. Zirkl M., Sawatdee A., Helbig U., Krause M., Scheipl G., Kraker E.*, et al.* An All-Printed Ferroelectric Active Matrix Sensor Network Based on Only Five Functional Materials Forming a Touchless Control Interface. *Adv. Mater.* **23,** 2069-2074 (2011).
4. Hu Z., Tian M., Nysten B. & Jonas A. M. Regular arrays of highly ordered ferroelectric polymer nanostructures for non-volatile low-voltage memories. *Nat. Mater.* **8,** 62-67 (2009).
5. Fukuda K., Sekitani T., Zschieschang U., Klauk H., Kuribara K., Yokota T.*, et al.* A 4 V Operation, Flexible Braille Display Using Organic Transistors, Carbon Nanotube Actuators, and Organic Static Random-Access Memory. *Adv. Funct. Mater.* **21,** 4019-4027 (2011).
6. Jadidian B., Hagh N. M., Winder A. A. & Safari A. 25 MHz ultrasonic transducers with lead- free piezoceramic, 1-3 PZT fiber-epoxy composite, and PVDF polymer active elements. *IEEE Trans. Sonics Ultrason.* **56,** (2009).
7. Zhao D., Katsouras I., Asadi K., Groen W. A., Blom P. W. M. & de Leeuw D. M. Retention of intermediate polarization states in ferroelectric materials enabling memories for multi-bit data storage. *Appl Phys Lett* **108,** (2016).
8. Lovinger A. J. Ferroelectric Polymers. *Science* **220,** 1115-1121 (1983).
9. Lovinger A. J. Poly(Vinylidene Fluoride). In: Bassett DC (ed). *Developments in Crystalline Polymers—1*. Springer Netherlands: Dordrecht, 1982, pp 195-273.
10. Bachmann M. A.&Lando J. B. A reexamination of the crystal structure of phase II of poly(vinylidene fluoride). *Macromolecules* **14,** 40-46 (1981).
11. Lando J. B., Olf H. G. & Peterlin A. Nuclear magnetic resonance and x-ray determination of the structure of poly(vinylidene fluoride). *J. Polym. Sci. A-1: Polym. Chem.* **4,** 941-951 (1966).
12. Gal´perin Y. L., Strogalin Y. V. & Mlenik M. P. *Vysokomol. Soed.* **7,** 933 (1965).
13. Martins P., Lopes A. C. & Lanceros-Mendez S. Electroactive phases of poly(vinylidene fluoride): Determination, processing and applications. *Prog. Polym. Sci.* **39,** 683-706 (2014).
14. Lovinger A. J. Annealing of poly(vinylidene fluoride) and formation of a fifth phase. *Macromolecules* **15,** 40-44 (1982).
15. Bachmann M., Gordon W. L., Weinhold S. & Lando J. B. The crystal structure of phase IV of poly(vinylidene fluoride). *J. Appl. Phys.* **51,** 5095-5099 (1980).
16. Davis G. T., McKinney J. E., Broadhurst M. G. & Roth S. C. Electric-field-induced phase changes in poly(vinylidene fluoride). *J.Appl. Phys.* **49,** (1978).
17. Davis G. T. & Singh H. Evidence for a new crystal phase in conventionally poled samples of poly(vinylidene fluoride) in crystal form II. *Polymer* **20,** (1979).
18. Naegele D., Yoon D. Y. & Broadhurst M. G. Formation of a New Crystal Form ($\alpha_p$) of Poly(vinylidene fluoride) under Electric Field. *Macromolecules* **11,** 1297-1298 (1978).
19. Li M., Wondergem H. J., Spijkman M.-J., Asadi K., Katsouras I., Blom P. W. M.*, et al.* Revisiting the δ-phase of poly(vinylidene fluoride) for solution-processed ferroelectric thin films. *Nat. Mater.* **12,** 433-438 (2013).

20. Garcia-Gutierrez M.-C., Linares A., Hernandez J. J., Rueda D. R., Ezquerra T. A., Poza P., et al. Confinement-Induced One-Dimensional Ferroelectric Polymer Arrays. *Nano Letters* **10,** 1472-1476 (2010).
21. Kang S. J., Bae I., Choi J.-H., Park Y. J., Jo P. S., Kim Y., et al. Fabrication of micropatterned ferroelectric gamma poly(vinylidene fluoride) film for non-volatile polymer memory. *J. Mater. Chem.* **21,** 3619-3624 (2011).
22. Ohigashi H. Electromechanical properties of polarized polyvinylidene fluoride films as studied by the piezoelectric resonance method. *J. Appl. Phys.* **47,** 949-955 (1976).
23. Ebnesajjad S. *Fluoroplastics volume 1: Non-Melt Processible Fluoroplastics*, 2000.
24. Baklar M. A., Koch F., Kumar A., Domingo E. B., Campoy-Quiles M., Feldman K., et al. Solid-State Processing of Organic Semiconductors. *Adv. Mater.* **22,** 3942-3947 (2010).
25. Collison D. *Methods in Rock Magnetism and Palaeomagnetism: Techniques and Instrumentation* Springer: The Netherlands, 2013.
26. Nakagawa K. & Ishida Y. Annealing effects in poly(vinylidene fluoride) as revealed by specific volume measurements, differential scanning calorimetry, and electron microscopy. *J. Polym. Sci.: Polym. Phys. Ed.* **11,** 2153-2171 (1973).
27. Dvey-Aharon H., Taylor P. L. & J. H. A. Dynamics of the field-induced transition to the polar α phase of poly(vinylidene fluoride). *J. Appl. Phys.* **51,** (1980).
28. Lovinger A. J. Molecular mechanism for a>d transformation in electrically poled poly(vinylidene fluoride). *Macromolecules* **14,** 225-227 (1981).
29. Sasabe H., Saito S., Asahina M. &Kakutani H. Dielectric relaxations in poly(vinylidene fluoride). *J. Polym. Sci. . A-2: Polym. Phys.* **7,** 1405-1414 (1969).
30. Miyamoto Y., Miyaji H. & Asai K. Anisotropy of dielectric relaxation in crystal form II of poly(vinylidene fluoride). *J. Polym. Sci.: Polym. Phys. Ed.* **18,** 597-606 (1980).
31. Miyamoto Y. Dielectric relaxation and the molecular motion of poly(vinylidene fluoride) crystal form II under high pressure. *Polymer* **25,** (1984).
32. Servet B.&Rault J. Polymorphism of poly(vinylidene fluoride) induced by poling and annealing. *J. Phys. France* **40,** 1145-1148 (1979).